\documentclass[11pt,a4paper]{article}
\usepackage[utf8]{inputenc}
\usepackage[T1]{fontenc}
\usepackage{amsmath,amssymb}
\usepackage[round,authoryear]{natbib}
\usepackage{graphicx}
\usepackage{booktabs}
\usepackage{tabularx}
\usepackage{array}
\usepackage{longtable}
\usepackage{hyperref}
\usepackage{url}
\usepackage{listings}
\usepackage{xcolor}
\usepackage{microtype}
\usepackage{parskip}

\newcolumntype{Y}{>{\raggedright\arraybackslash}X}
\newcolumntype{L}[1]{>{\raggedright\arraybackslash}p{#1}}
\newcolumntype{C}[1]{>{\centering\arraybackslash}p{#1}}

\hypersetup{
    colorlinks=true,
    linkcolor=blue,
    citecolor=blue,
    urlcolor=blue,
    pdftitle={When Does an Interpretation Count as Established?},
    pdfauthor={Deyu Jing},
}

\title{\textbf{When Does an Interpretation Count as Established?}\\[0.4em]
\large The Formation, Evaluation, and Responsibility of Interpretation in Generative AI}
\author{Deyu Jing\\
        Fudan University, Shanghai, China\\
        \texttt{baireinhold@163.com}}
\date{September 4, 2026 \quad Version 1.5}

\begin{document}

\maketitle

\begin{abstract}

Generative AI research increasingly evaluates factuality, citation, coverage, and report structure. Yet passing these local checks does not by itself show that a humanistic interpretation has been established. This paper asks how an interpretation comes to be recognized within sociotechnical processes. It develops three connected concepts. Interpretive appearance names the gap between the finished form of an output and the publicly traceable process through which materials, counterevidence, and revisions constrained the judgment. The evaluation contract names the bounded materials, tasks, criteria, permitted inferences, and failure conditions within which a local judgment is valid. Standing substitution names the unwarranted conversion of a genuine local pass into a stronger claim that an interpretation, result, or research capability has been established, without commensurate new evidence or bridging arguments. The paper then examines responsibility for judgment: a text may acquire recognition while no public structure remains for stating reasons, answering objections, revising, or withdrawing the conclusion. Humanistic scholarship is a revealing test because new materials and conceptual distinctions can alter both the question and the criteria of evaluation. The paper develops delayed closure as a practice of keeping recognized interpretations revisable and proposes five public requirements concerning materials, evidence, failure, revision, and responsibility. The argument is conceptual and normative. It does not offer a benchmark or determine whether models possess understanding. It explains why local evaluation, finished textual form, and public recognition are insufficient evidence that an interpretation has been formed.

\end{abstract}


\section{Introduction --- When Does Interpretive Appearance Cease to Suffice as Evidence That a Judgment Has Been Formed}\label{introduction-when-does-interpretive-appearance-cease-to-suffice-as-evidence-that-a-judgment-has-been-formed}

In competitive sports, spectators ordinarily do not conclude from a single brilliant move that an athlete already possesses a stable ability. A performance can count as evidence of ability because it is usually connected, in a traceable way, to training, failure, rules, and testing by a community; this connection does not guarantee that every performance is correct, but it allows the performance to be placed within a longer relation of judgment. Generative artificial intelligence is changing the visibility of this connection: finished text can circulate before the process of its formation has been accounted for and before the boundaries of its evaluation have been preserved.

Current academic controversies over generative AI have formed two important but individually limited paths. One asks whether language models understand, whether they have intentions, or whether they can bear responsibility. The other decomposes outputs into facts, sources, verifiability, coverage, reasoning, and report structure, and confirms local quality through finer-grained evaluation. The latter path has made it possible to judge long-form research outputs by more than overall impression, and has made it possible to locate errors, mismatched citations, and insufficient support. Yet even when an output passes established checks of factuality, citation, coverage, and logic, one question does not thereby disappear: when are these local passes sufficient for a community to recognize that an interpretation has been formed?

The question can be presented through a minimal counterexample that does not depend on the inner states of any model. Suppose that every citation in a historical report is genuine, that every page can be linked back to its source, that factual verification and source attribution have both passed, and that the report's structure satisfies the required claim--support relations; but the report's key causal judgment merely repeats a piece of secondary scholarship, without reworking the counterevidence in the primary materials and without explaining why the secondary author's generalization applies to the present problem. If this report is subsequently named by papers, platforms, or institutional registries as ``a historical interpretation that has been formed,'' its defect cannot be exhausted by the categories of ``hallucination'' or ``citation error.'' The local evaluations may be genuine; the problem occurs where a local pass is endowed with a stronger interpretive standing.

This paper accordingly poses a narrower research question: how does an interpretation, presented in the appearance of a finished product, acquire the standing of being ``already established'' in sociotechnical circulation, and how can this standing come apart from conditions of formation, evaluation boundaries, and responsibility for judgment? Interpretive standing here is not a synonym for interpretive truth, nor is it an inference about whether an author inwardly read, reflected, or revised. What this paper examines are public conditions: whether readers can return to the materials and versions; whether they can distinguish facts from inferences; whether they can see how counterexamples changed the judgment; whether they can know under what circumstances a conclusion should be downgraded or withdrawn; and whether there exists an agent or structure that can give reasons and bear consequences.

The argument of this paper proceeds on three interconnected but mutually irreplaceable levels. First, building on Adhya's pedagogical observation of ``interpretive appearance'' in the classroom, we translate it into a question of formation at the level of the research community, and we define interpretive appearance as the divergence between the finished appearance of the output and publicly traceable conditions of formation. This concept does not simply relabel fluent text as mere appearance, nor does it claim that all research using generative systems lacks a process of formation; it examines only whether, when a text asks its readers to believe that an interpretation has been settled, the public record provides conditions of formation commensurate with that demand.

Second, this paper proposes the working term ``evaluation contract,'' which refers to the scope and basis of an evaluation. It explains why local evaluation must be provisionally closed within bounded materials, tasks, scoring, permitted inferences, and failure conditions, and it proposes ``standing substitution'' to describe a specific conversion from one judgment to another: a genuine local pass is treated by later papers, platforms, or dissemination materials as sufficient warrant for a stronger claim of interpretation or achievement, without commensurate new evidence or bridging arguments. Treating a local pass as if the whole were established is not a rejection of evaluation, nor another name for inadequate measurement in general; it requires that the original evaluation be genuine and valid, that the subsequent claim be stronger, that the original evaluation be used as the key warrant, that no commensurate new support has been added, and that the original boundaries and conditions for downgrading no longer constrain the subsequent, stronger claim.

Third, this paper turns the question of responsibility away from the ontological controversy over ``whether machines can be responsible'' and back toward the public practical relations of judgment. A generated text may acquire recognition while the judgment as a whole still has no identifiable attribution, no statable reasons, no procedure by which counterexamples can change the judgment, no executable path of downgrading or withdrawal, and no person or explicit structure that bears the consequences. The ``delayed closure'' proposed on this basis is not a matter of the slower the better; it means postponing the moment at which an interpretation is made irrevocable, and keeping the record of formation, the scope and basis of evaluation, and the relations of responsibility open to re-examination, so that revision and withdrawal remain possible. Section 6 will translate this requirement into five public conditions, while acknowledging that they only increase the likelihood of discovery, contestation, correction, and withdrawal, and do not guarantee that an interpretation is true.

The paper unfolds as follows. Section 1 reviews the progression from the stochastic-parrots critique, through hallucination and source attribution, to the evaluation of long-form research, and argues that local verifiability is a necessary floor for the recognition of an interpretation, not a sufficient condition. Section 2 poses the problem of formation and delimits it against hallucination, plausibility, trustworthy interfaces, capability inference, and illusions of understanding. Section 3 defines the scope and basis of evaluation and discusses warranted elevation of standing together with two strict counterexamples, in order to show the boundaries of ``treating a local pass as the whole being established.'' Section 4 turns to responsibility, reconstructing responsibility for judgment through footnotes, context, formation, and practical standards, while keeping its distance from neighboring problems such as the responsibility gap. Section 5 takes humanistic scholarship as the testing ground, shows how materials and counterexamples can rearrange questions and criteria, and defines delayed closure. Section 6 then converts the analysis of these three levels into several concrete requirements: that one can return to the original materials and the original judgment, that different evidential levels can be distinguished, that failure can be acknowledged, that revision is possible, and that a bearer of responsibility can be found.

To avoid conceptual conflation, this paper draws the following distinctions. ``Cross-level translation'' is the neutral process by which an evaluation result enters a new vehicle; ``warranted elevation of standing'' means that a subsequent, stronger claim has acquired commensurate new materials, independent review, and bridging arguments; ``standing substitution'' is the improper form of this process. What this paper means by an interpretation being recognized is, at a minimum, that some vehicle permits an output to serve as an acceptable premise for subsequent judgment, evaluation, citation, teaching, resource allocation, or public knowledge activity. This does not mean that the interpretation is true, nor does it require that the community has reached a final consensus.

\section{From ``Stochastic Parrots'' to ``An Interpretation Has Been Established''}\label{from-stochastic-parrots-to-an-interpretation-has-been-established}

\subsection{A Question That Has Already Been Advanced}\label{a-question-that-has-already-been-advanced}

Academic controversy over generative AI often begins with a question that is at once forceful and easily closed: does a language model actually understand the text it generates? The value of the ``stochastic parrots'' critique lies in its refusal to infer meaning, communicative intent, and the capacity for responsibility directly from formal linguistic success. A model can generate coherent, grammatical, prompt-responsive text on the basis of statistical relations in its training corpus, but this formal success by itself cannot show that the text bears the meaning carried by human utterance, nor can it show who should be responsible for the judgments it contains. Bender and her co-authors (2021) at the same time direct attention to training materials, deployment environments, and relations of power: how meaning, knowledge, and responsibility come to be ascribed to machine outputs depends on a set of sociotechnical conditions, not merely on the scale of a model's parameters.

This critique remains necessary, but it is insufficient for dealing with today's long-form research outputs. If the question remains ``does the model understand,'' all research risks are easily compressed into whether some capability exists inside the machine; if the question is changed to ``does the text contain errors,'' the selection of materials, the relations among sources, and the inferential processes that occur before an interpretive judgment is decomposed are hidden from view. This paper starts from a fact that technical research has already established: the statements, sources, and long-form organization of generative systems can each be evaluated separately. The question thereby shifts from whether the model possesses meaning to what entitles a finished text to be recognized.

\subsection{From Hallucination to Sources and Support Relations}\label{from-hallucination-to-sources-and-support-relations}

The first object to come into view was hallucination. Hallucination research singles out the relation between a statement and external facts, requiring that one judge whether a sentence can be supported by available evidence; source-attribution research goes further, asking whether the listed literature actually exists, whether it in fact supports the statement, and whether the citation is placed in a position commensurate with its role. That a page really exists does not mean that the page supports every judgment attached to it; that a statement is broadly correct does not mean that its chain of sources is complete. Retrieval-augmented generation and the corresponding RAG evaluations place retrieved materials, candidate evidence, and generated statements within a single framework of checking, so that omissions, mismatches, and insufficient support can be identified item by item.

This work has changed the meaning of ``factuality.'' A fact is no longer a reader's overall impression of an article; one must first decompose the statements, confirm the role of each piece of evidence, and then judge whether, within the given range of materials, the support is sufficient. The relational checks established by ALCE (Gao et al.~2023), RAGChecker (Ru et al.~2024), and research on source fidelity (Rashkin et al.~2023) can show that an answer's citations fail to support the corresponding claims, or that the retrieved context omits decisive materials. They therefore constitute an indispensable foundation for research with generative systems. Yet the verifiability of source relations still only answers the question ``what is the relation between this sentence and these materials''; it has not answered ``why was the question formed in this way,'' ``which conceptual distinctions changed the judgment,'' or ``did competing interpretations have a chance to defeat the original conclusion.''

\subsection{Long-Form Text Makes Local Strengths and Weaknesses Easier to See}\label{long-form-text-makes-local-strengths-and-weaknesses-easier-to-see}

As systems began to generate long-form reports, a single rate of factual correctness could no longer bear the entire burden of evaluation. LongFact decomposes long answers into independently verifiable factual claims, SAFE checks these claims through search and item-by-item inspection, and VeriScore takes verifiable claims as the unit of analysis (Wei et al.~2024; Song et al.~2024). These methods do not claim that a report consists only of facts; their role is to separate out, in advance, the checkable relations otherwise blended into fluent narration, so that researchers know which sentences are stating facts and which are advancing inferences that go beyond the facts.

Coverage and report structure thereby became another set of objects of evaluation. ICAT attends to whether an answer addresses the aspects prescribed in advance by the task (Samarinas et al.~2025); DeepResearch Bench goes further, incorporating source quality, synthesis of materials, causal connection, multiple perspectives, comparative explanation, and insight into expert criteria (Du et al.~2025). ResearchRubrics acknowledges that deep-research answers may have multiple valid paths, does not require a system to reproduce a single reference text, and instead uses fine-grained criteria to judge whether factual grounding, reasoning, and expression meet the requirements (Sharma et al.~2025). \emph{ReportLogic} takes as its object of judgment whether readers can follow a report's analytic arc, its context, its claim--support relations, and its bridging reasons, and it uses adversarial examples to show that adding headings, numbering, summaries, or ``evidence'' labels may improve surface organization without necessarily adding underlying support (Zhao et al.~2026).

It is important to acknowledge that this work carries real information. It has neither crudely reduced complex research to error-checking nor abandoned evaluation because open-ended tasks are hard to exhaust; it can show on which dimensions of factuality, sourcing, coverage, reasoning, or expression a system has improved, and it can help researchers compare the local differences among systems on the same task. The evaluation contract discussed later presupposes precisely that such local evaluations can hold. The question that remains to be added is this: once a local evaluation has been confirmed, under what conditions is the result understood as an interpretation that has been formed?

\subsection{Open-Ended Tasks Also Need Their Scope Stated First}\label{open-ended-tasks-also-need-their-scope-stated-first}

DeepResearch Bench II displays this duality in open-ended tasks most clearly (Li et al.~2026). It extracts facts and inferences from survey articles written by experts and, through automatic extraction, manual cleaning, and revision by domain experts, forms fine-grained evaluation criteria; some of the expert articles are designated as shielded sources, and the system must reconstruct their content from other materials. Such a design allows answers to remain open while still specifying which facts must appear, which inferences count as non-trivial analysis, and which sources may not be reproduced directly. What the system is asked first is whether, given the question, the materials, and the criteria, it can find the necessary evidence and establish the corresponding relations; it has not yet been asked whether it can re-specify the question, change the identity of the evidence, or propose a different scope and basis of evaluation.

This is not an accusation against the benchmark but an accurate description of what it measures. The expert articles and fine-grained criteria make comparison possible and provide localizable failure information for system improvement; at the same time, they provisionally convert openness into a set of checkable task conditions. ResearchRubrics permits multiple valid paths, \emph{ReportLogic} attends to the internal relations of a report, and DeepResearch Bench II prevents simple reproduction by shielding sources (Sharma et al.~2025; Zhao et al.~2026; Li et al.~2026); the differences among the three show precisely that ``openness'' does not mean the absence of boundaries. An answer can display very high quality within a task and still have obtained only a local pass within that task.

\subsection{Local Evaluation Cannot by Itself Show That an Interpretation Is Established}\label{local-evaluation-cannot-by-itself-show-that-an-interpretation-is-established}

From the discussions of construct validity and benchmark design, any score must answer two questions: what exactly it measures, and which interpretation and which decision it is used to support. The work of Raji and her co-authors on benchmark audits (Raji et al.~2021) reminds us that an evaluative construct cannot be detached from the goals of the task and the actual scenarios of use; Schlangen's critique of language evaluation (Schlangen 2023) likewise points out that the comparability of scores is not equivalent to the construct having been adequately confirmed. This paper accepts this methodological vigilance: it neither writes local evaluation off as a meaningless proxy, nor writes open interpretation off as arbitrary activity without standards.

Likewise, the neighboring research on Proxy-Based Evaluation has made it clear that performance metrics, benchmarks, transparency documentation, and auditing procedures are all proxy evidence for governance goals; even when the proxies are satisfied, distributional shifts in deployment, hidden harms, and runtime failures may remain undiscovered. This paper follows the demarcation that ``a proxy is not the target'' and goes on to trace a more specific cross-vehicle question: after a proxy result leaves its task, is it rewritten by papers, platforms, registries, or dissemination materials into ``the interpretation has been established'' or ``the system already possesses general research capability''? Within the original evaluation scope, a local pass may be entirely genuine; the problem occurs when the result is endowed with a larger identity without the addition of commensurate materials and bridging reasons.

\subsection{The Difference Between Local Testability and Overall Interpretation}\label{the-difference-between-local-testability-and-overall-interpretation}

The current literature has therefore established a floor that cannot be bypassed: without minimal testability of facts, sources, coverage, and report logic, no research output should ask a community to treat it as an interpretation. But this floor is not a sufficient condition. A report can pass every prescribed item of statement, citation, coverage, and structure and still have completed only a checkable answer within a bounded task; it may not have explained why the materials belong to the relevant scope, may not have engaged evidence that runs against its main line, and may not have displayed the bridge from local relations to an overall interpretation. If these limits are preserved, a local pass can function accurately; if they are omitted in subsequent circulation, the finished appearance of the output may acquire standing ahead of the process of its formation.

The research object of this paper is thereby confined to a stretch of relations that technical evaluation has opened but not yet closed: rather than inventing a replacement concept for meaning, factuality, sourcing, or report logic, it examines how these local relations are organized, named, and carried into a community's interpretive judgment. The next section will define the divergence between the finished appearance of the output and the process of its formation as ``interpretive appearance,'' and will explain how it differs from hallucination, plausibility, trustworthy interfaces, and illusions of understanding. Only after this formation layer has been distinguished can one further ask how the evaluation contract translates a local pass into interpretive standing.

\section{Why an Interpretation Comes to Look Finished}\label{why-an-interpretation-comes-to-look-finished}

\subsection{From ``Looks Like an Interpretation'' in the Classroom to Recognition in Research}\label{from-looks-like-an-interpretation-in-the-classroom-to-recognition-in-research}

Raahi Adhya's (2026) observation of the humanities classroom provides this section with a starting point that must be acknowledged. What she describes is not a general decline in textual quality but a more specific misalignment: generative AI can produce fluent, orderly analytic prose with a critical tone, even replicating the surfaces of close reading, contextualization, and theoretical judgment, yet this prose does not necessarily issue from processes of reading, discussion, and revision commensurate with its degree of finish. She summarizes this change as a shift of learning from process to product, and she notes that the errors, hesitations, drafts, and discussions of in situ writing sometimes preserve traces of formation. The ``sometimes'' here is crucial: errors are not sufficient proof of understanding, and drafts can be forged; they are worth noticing only because they may keep a judgment in a state in which it can still be questioned and changed.

This paper takes up this observation (Adhya 2026), but it does not enlarge a problem of classroom learning directly into a moral judgment on the research community. What Adhya faces is how a teacher, in evaluating a student's assignment, can tell whether learning has occurred; what this paper asks is on what basis a community, when a piece of interpretive prose leaves the classroom and enters papers, databases, platforms, or public circulation, recognizes it as ``an interpretation that has been formed.'' There is a change of scale between the two: classroom evaluation can require students to account for their process, whereas research results usually circulate through different institutions as finished products, citations, and signatures. It is precisely in this circulation that conditions of formation may be compressed while the finished appearance of the output is preserved, so that the text acquires an interpretive standing that exceeds its original evidential record.

This distinction also delimits the relation between this paper and the stochastic-parrots critique. Bender et al.~(2021) remind us that formal linguistic fluency cannot directly demonstrate understanding of meaning, communicative intent, or the capacity for responsibility; this reminder remains valid, but it does not tell us when a text already written in the form of an interpretation acquires institutional status within a community. This paper need not first determine whether a model understands, nor prove that a user has not understood, in order to pose a narrower public question: whether the publicly available materials are sufficient to support the degree of finish a text presents, and whether readers can trace along the record how the judgment was constrained by materials and counterexamples.

\subsection{The Interpretation Looks Finished, but Its Formation Has Not Been Accounted For}\label{the-interpretation-looks-finished-but-its-formation-has-not-been-accounted-for}

This paper calls this misalignment in public conditions ``interpretive appearance.'' It is not the aesthetic impression of ``looking like an interpretation,'' but a relational judgment about standing:

\begin{quote}
Interpretive appearance arises when an output presents the form of a completed interpretation, while the publicly available record is insufficient to trace how materials, counterevidence, and revisions constrained that judgment, or is insufficient to reopen that judgment for examination by the community.
\end{quote}

The ``conditions of formation'' in the definition refer, first of all, to the epistemic records relevant to interpretive standing that others can return to: the scope and versions of the materials, the roles that sources play in the argument, the bridging inferences connecting facts to concepts, the counterexamples considered or excluded, the versions in which the judgment changed, and the conditions under which the conclusion should be downgraded or withdrawn.\footnote{``Conditions of formation'' do not require disclosing all prompts, private notes, or chains of thought, and do not presuppose inferences about an author's psychological processes.} Public traceability requires that the interpretive credit carried by a finished product cannot be wholly detached from these minimal conditions.

Code provides a boundary case worth pausing over here. For local tasks with explicit specifications, relatively stable inputs and environments, and executable tests, code has a comparatively strong ``artifact--function'' coupling: compilation, execution, and testing can publicly confirm, fairly quickly, whether a piece of implementation produces the expected behavior. But this coupling cannot be rewritten as a unity of ``artifact--formation of judgment.'' What runtime verifies is the behavior of the artifact within a specific evaluation scope; it does not verify that the requirements were specified correctly, that the architectural trade-offs were reasonable, that untested conditions are safe, or that anyone actually understood the code.

If a person requests only an explicit function and limits the claim to the test results, they have not thereby advanced any claim to programmer standing that exceeds the record; if a successful run is subsequently written up as the problem having been correctly understood, the system being ready for production, or AI having acquired a stable software-engineering capability, the problem reappears at a higher level. Code thus shows that interpretive appearance attaches to claims of standing, not to the material properties of a kind of text or artifact.

``Divergence'' does not mean that a process of formation does not in fact exist. A researcher using a generative system may have read, verified, and rewritten at length without preserving these relations in the final product; conversely, a text written entirely by a human may manufacture the appearance of completion through terminology, citations, and orderly structure alone. This paper therefore makes no judgment about the inner states of authors; it examines only whether there is a connection, commensurate with the strength of the claim, between the finished appearance of the output and the public record. This qualification allows the paper to acknowledge genuine human--machine collaboration and genuine processual labor at the same time, and to point out that if such labor cannot leave, at the key judgments, conditions that allow one to return to the original materials and the original judgment, the recognition of an interpretation may still circulate ahead of the process of its formation.

\subsection{It Is Not Hallucination}\label{it-is-not-hallucination}

The nearest neighbor most likely to cause confusion is ``hallucination.'' Hallucination research asks whether statements accord with external facts, and source-attribution research asks whether the listed literature exists and supports the corresponding claims; these relations cannot be omitted from any research output. But the situation described here can arise without any fabricated facts. A report may cite real books, and the pages may indeed contain the quoted sentences, yet the report may have selected only materials supporting a particular narrative, without explaining why contrary materials do not change the judgment, or it may directly elevate a generalization from secondary scholarship into an interpretation of the historical object. The problem here is not that ``the citation does not exist,'' but that the finished appearance of the output implies that the materials have been adequately examined, while the public record is insufficient to support that implication.

Likewise, retrieval-augmented generation, item-by-item factual verification, and verifiable citations can reduce errors, but they do not automatically eliminate interpretive appearance. They increase the checkability of the relations between statements and sources and between statements and facts, but they do not necessarily explain why a question was posed in this way, how a conceptual distinction was formed, or whether competing interpretations were genuinely engaged. So long as summaries, candidate leads, and local analyses retain their limited identities, they do not constitute the formation layer; what is needed here is a stronger kind of presentation, namely that the text leads readers to believe the interpretation has been settled while the material conditions for re-examining it have not correspondingly appeared.

\subsection{It Is Not Plausibility or a Trustworthy Interface Either}\label{it-is-not-plausibility-or-a-trustworthy-interface-either}

Another group of neighboring research attends to why texts are easily believed. Discussions of AI-generated plausibility point out that fluency, coherence, and familiar structure in language may be mistaken as grounds for knowledge, and they distinguish this transfer --- from plausibility cues, to ascribed reasons, to institutional validation --- from appropriate trust, testimony, and automation bias. The research of Liu and her co-authors on the verifiability of generative search engines (Liu et al.~2023) uses ``trustworthy appearance'' to describe the phenomenon in which results carry inline citations without adequately supporting the attached claims. This work accurately explains why readers accept an insufficient result.

This paper pushes the question one level further, but it does not rename this work as the formation layer. Plausibility is the cue by which a text is accepted; a trustworthy interface is the way in which source relations are presented; interpretive appearance attends to how these cues combine with the finished appearance of an interpretive product and acquire the standing of ``already formed'' within a community. A text may have no conspicuous interface design and yet, through paragraph structure, conceptual terminology, and the tone of its conclusions, lead readers to treat candidate relations as established interpretation; it may also have an excellent citation interface and still not show how counterexamples and bridging inferences changed the judgment. What the formation layer adds is not a psychological description of ``belief,'' but a tracking of the conditions of standing.

\subsection{The Distance from Two Kinds of ``Illusion of Understanding''}\label{the-distance-from-two-kinds-of-illusion-of-understanding}

Research on model capability over-inference warns against inferring broader understanding from a model's local performance on tests: a model may display understanding-like performance on certain prompts, tasks, or surface behaviors, but this performance does not automatically support strong claims about internal capabilities. Messeri and Crockett's (2024) account of ``illusions of understanding'' in scientific research places the focus on the researcher's side, pointing out that AI tools' promises of productivity and objectivity may exploit human cognitive limits, leading people to believe they understand more of the world while obscuring the formation of methodological, topical, and perspectival monocultures.

The two address capability inference and researchers' epistemic illusion respectively; neither is the object of this paper. This paper neither judges whether a model possesses some hidden capability nor measures whether researchers actually feel they understand more; it analyzes only whether a text that has entered academic procedures has acquired an interpretive standing that exceeds the public record of its formation. Even if readers are fully aware that a model merely predicts text, and even if no one suffers any subjective illusion, institutions may still enter a formally complete text under the heading of ``interpretive achievement'' or ``research capability.'' Conversely, even if a researcher believes their understanding is adequate, so long as the output is explicitly marked as a candidate lead and material boundaries, counterexamples, and conditions for withdrawal are preserved, one cannot diagnose the formation layer on that basis alone.

\subsection{Two Basic Conditions and Four Cases That Should Not Be Conflated}\label{two-basic-conditions-and-four-cases-that-should-not-be-conflated}

To avoid sweeping every finished product into interpretive appearance, at least two thresholds must be passed. The first threshold is that the text must put forward interpretive statements, rather than merely transcribing facts, listing summaries, generating keywords, or offering candidate associations still awaiting verification. Interpretive statements attempt to articulate relations of meaning among materials, conceptual structures, historical positioning, or causal connections, and they therefore carry a heavier burden of reasons than the transportation of information. The second threshold is that the finished appearance of the output must exceed what the public evidential conditions can directly support: the text does not merely say ``here is a possible direction,'' but presents itself, in the tone of a conclusion, a synthetic judgment, or a scholarly achievement, as having examined the materials, weighed the counterexamples, and completed closure.

Four negative boundaries follow. First, summaries, indexes, candidate leads, and local analyses do not constitute interpretive appearance so long as they retain their limited identities; they may be useful, but they have not claimed that an interpretation has been established. Second, completing a bounded task under a contract does not amount to appearance; if the task explicitly specifies the scope of materials, the use, and the failure conditions, and the result never exceeds these boundaries, contractualization can instead make standing more accurate. Third, genuine citations, linkable pages, logs, and version records are not a sufficient guarantee; they become part of the conditions of formation only when they can change the evaluation of the overall judgment. Fourth, the formation layer does not require disclosure of the entire process, and it does not treat slowness, errors, hesitation, or a purely human signature as marks of authenticity. An efficient and refined research process can leave adequate public reasons; a pile of drafts may carry no interpretive significance at all.

These boundaries separate the formation layer from general criticisms of ``excessive finish.'' The problem is not that the text is well written, nor that the model writes like an expert, but that the conditions of formation demanded by the finished appearance of the output have not been correspondingly borne by the public record. If a system's report explicitly states that the materials have not been exhausted, that the conclusions serve only for topic selection, and that counterexamples have not been retrieved, and if it hands its results back to researchers for renewed judgment, then it may be fluent without having acquired interpretive appearance. The problem is activated only when the same report is rewritten as ``research has discovered a certain historical mechanism'' while the original record has gained no new materials or bridging reasons.

Interpretive appearance must also be distinguished from interpretive standing itself. What this paper means by interpretive standing is the institutional status of a text being provisionally recognized by a community as entitled to put forward an interpretation; interpretive authority concerns how much trust such an interpretation should receive on a given occasion, and legitimacy further concerns whether it conforms to some norm or procedure. The three may appear together, or they may be misaligned with one another. A text marked as a ``candidate interpretation'' may have great research value without yet having acquired stable interpretive standing; an already published interpretation may possess standing and authority yet lose its legitimacy in the face of new materials, or require withdrawal. This paper diagnoses only the relation, at the moment standing is acquired, between the finished appearance of the output and the conditions of formation; it does not prejudge whether the interpretation is correct, whether the authority is justified, and it does not equate the absence of a public record directly with institutional violation. Precisely because standing, authority, and legitimacy are not the same thing, the later chapters must ask separately how local evaluations raise standing and who bears the overall judgment once standing has been acquired.

Nor are conditions of formation a single level of transparency identical for all claims. A bounded statement about a single fact may need only preserved versions and sources; an interpretation of conceptual change must additionally account for context, senses of words, competing materials, and the bridge from local evidence to historical generalization. The same record can therefore suffice to support a weaker candidate judgment while being insufficient for a stronger causal or global judgment. Judgments at the formation layer must vary with the strength of the claim, the type of materials, and the occasion of use; one cannot treat ``having disclosed some process information'' as a universal pass. Conditions of formation can also be preserved jointly by multiple agents: the researcher's version notes, an archive's access records, reviewers' objections, a team's division-of-labor table, and a platform's withdrawal mechanism may each bear part of the function. What matters is not whether these records are concentrated in a single file, but whether they can be interconnected, so that readers know which piece of material constrained which judgment, and what change would disqualify the original judgment. If the records are isolated from one another, they can only prove that the system once ran; they cannot show how the materials changed the judgment, and the finished appearance of the output may still acquire interpretive status ahead of schedule.

\subsection{From ``Looking Finished'' to Local Evaluation}\label{from-looking-finished-to-local-evaluation}

This section has only described the relation between the finished appearance of the output and the public conditions of its formation; it has not yet explained why this form can enter papers, platforms, or institutional procedures and acquire higher status. An interpretive appearance can remain in a personal notebook, and it can be rejected by readers; it acquires practical effect only when it is treated, within some chain of evaluation and dissemination, as a result that has ``already been formed.'' The formation layer therefore cannot by itself yield standing substitution, still less the absence of responsibility.

The next section turns the question toward the evaluation contract: in order to judge facts, sources, coverage, and report logic, evaluators must provisionally fix materials, tasks, criteria, and permitted inferences; these fixings make local comparison possible, but they also provide the entry point for the subsequent cross-level translation of results. What needs to be asked is not whether evaluation is useful, but, when a local pass leaves its original contract, how the finished appearance of the output is endowed with a stronger interpretive standing, and whether the original material boundaries and failure conditions continue to travel with the result.

\section{When Does a Local Evaluation Get Taken as the Whole Being Established}\label{when-does-a-local-evaluation-get-taken-as-the-whole-being-established}

\subsection{Evaluation Is Not the Enemy of Interpretation}\label{evaluation-is-not-the-enemy-of-interpretation}

The long-form research outputs of generative AI have deprived an old dichotomy of its explanatory power: at one end, interpretation is reduced to whether the model truly understands; at the other, the evaluation of interpretation is reduced to whether the text contains factual errors. Recent research on source attribution, long-form factuality, citation verifiability, and report logic has shown that the relations between statements and facts, between statements and sources, and between report structure and support relations can be separated into distinct objects of evaluation. Evaluation is therefore not a crude substitute external to humanistic scholarship; it can reveal what a report has omitted, whether a source really supports the attached claim, and whether a causal connection lacks a bridging reason, and it can also display the local differences among systems on the same task.

The problem appears at the next step. An evaluation result usually has to leave a single run, a dataset, or a set of scoring criteria and enter papers, platforms, institutional registries, project descriptions, and dissemination materials. In the course of this movement, whether the material boundaries, task uses, permitted inferences, and failure conditions originally attached to the result remain visible is not automatically decided by the score. That a report is genuinely qualified on the present task cannot ground the conclusion that a stronger judgment has acquired the standing of open interpretation. To analyze this conversion, one must first explain how local evaluation can hold at all, and why it must be provisionally closed.

\subsection{What the Scope and Basis of Evaluation Is}\label{what-the-scope-and-basis-of-evaluation-is}

This paper calls the materials, tasks, scoring, permitted inferences, and failure conditions that enable local evaluation to operate an ``evaluation contract'': the evaluator must know which batch of materials to check, which question to address, which standards to judge by, what may be inferred from the result, and which circumstances should be recorded as partial satisfaction, insufficient evidence, or inability to answer.\footnote{``Contract'' is a working name: it constructs no legal relation and does not presuppose that evaluators, systems, and users have reached actual agreement; only the sense of ``a set of operative constraints'' is intended.}

Without such constraints, open-ended outputs cannot be stably compared. \emph{ReportLogic} places a given query, a retrieval context, and a candidate report within a single evaluation unit, checking whether the report forms a unified analytic arc, provides the necessary context, and establishes explicit relations between claims and support; it also uses adversarial examples to show that adding headings, numbering, lengthy explanations, or ``evidence'' labels may improve surface presentation without repairing the underlying support relations. Being qualified here is not a natural property of the report, but the result of the report within a specific query, context, set of dimensions, and judgment procedure. This boundary does not diminish the value of \emph{ReportLogic}; on the contrary, it enables \emph{ReportLogic} to state clearly what it measures.

\texttt{ResearchRubrics} addresses another kind of openness. It acknowledges that deep-research answers are long, formally diverse, and may have multiple valid paths, and therefore does not require the system to reproduce a single complete reference text; instead it uses fine-grained criteria written by experts to check factual grounding, whether the reasoning holds, and clarity of expression. Its abstract reports that even the leading system's rate of criterion compliance remains below 68\%, with the main defects being the omission of implicit context and insufficient reasoning over retrieved information. This result shows that fine-grained criteria can yield genuine diagnostic information, and also that ``openness'' does not mean the absence of an evaluation contract: the answer need not be unique, but the scope of materials, the task goals, and the acceptable inferences are still specified in some way.

\texttt{DeepResearch\ Bench\ II} displays the pre-set conditions within open-ended tasks even more directly. The benchmark extracts facts and inferences from survey articles written by experts and, through automatic extraction, manual cleaning, and revision by domain experts, forms fine-grained evaluation criteria; some of the expert articles are listed as shielded sources, and the system must reconstruct their content from other materials. This does not treat the expert articles as eternally correct answers; it confirms, within a specified task, whether the system found the necessary evidence and produced the corresponding analysis. An open-ended task therefore still needs to specify provisionally what counts as a relevant fact, what counts as non-trivial analysis, and which sources may not be reproduced directly. The existence of a scope and basis of evaluation is a condition for comparison to begin, not a proof that interpretation has been completed.

Software testing makes the point about the ``evaluation contract'' especially clear. A test suite specifies bounded inputs, task goals, pass criteria, and failure conditions, so that ``the tests pass'' is, first of all, only a local pass of the artifact within that evaluation scope. It cannot by itself yield the conclusions that the system is globally correct, that the requirements have been correctly understood, that the code is safe enough to deploy, or that the coder should thereby be recognized as possessing a stable software-engineering capability.

The conversion that occurs here is structurally analogous to what this paper calls standing substitution: the original evaluation result is genuine, but the subsequent, stronger claim adds new burdens of specification interpretation, architectural judgment, untested conditions, maintenance, and responsibility. If these burdens are not added, testing turns from a useful local judgment into an unduly enlarged credential of standing.

\subsection{Provisionally Closed, but Open to Re-examination}\label{provisionally-closed-but-open-to-re-examination}

The ``provisional closure'' of the scope and basis of evaluation is necessary: if materials, questions, scoring, and failure records changed without limit in every judgment, local results would not be comparable. But the demand of open interpretation is not to remain forever inside the evaluation scope. New materials may make the original question no longer the most important one; counterexamples may invalidate the old evaluation criteria; differences among versions may change the identity of the evidence; competing interpretations may reject the original decomposition of the question. The scope and basis of evaluation must therefore be closed within a single act of judgment and yet remain open to re-examination within the community's controversies.

Validity theory can state this distinction more clearly. The Messick tradition of construct validity (Messick 1989) requires examining the interpretation that a measurement supports together with its social consequences; Kane's (2013) argument-based validity requires explicitly listing the inferences from observations to interpretation and use, together with their supporting assumptions.\footnote{Messick, Gadamer, and MacIntyre are used here only to delimit the argument's theoretical boundaries; they do not replace this paper's own conceptual work. Claims about AI evaluation are cited to the specific benchmark and evaluation studies listed in the references.} Kane emphasizes in particular that the object of validity is the proposed interpretation and use, not a test or a score detached from its use; more ambitious claims require more evidence, and interpretations and uses change with new needs and understandings. This paper accepts this work and does not write the ``evaluation contract'' as a replacement for validity theory.

What this paper adds lies at the two ends of the evaluation chain. At one end is how the contract, before evaluation, selects materials, organizes the question, and specifies the permitted inferences; at the other end is whether, after evaluation, when the result enters papers, platforms, registries, and dissemination, the original boundaries, uses, and conditions for downgrading continue to constrain subsequent claims. A validity argument can show how a score or a report is supported under a specific interpretation and use, but it does not automatically show whether, once the result enters a new vehicle, it has been exchanged for a larger identity. The latter requires tracking how wording, evidence, and responsibility move among vehicles.

\subsection{Treating a Local Pass as the Whole Being Established: Five Conditions}\label{treating-a-local-pass-as-the-whole-being-established-five-conditions}

This paper calls this specific conversion ``standing substitution,'' that is, treating a local pass as if the whole were established. More precisely, standing substitution is the unwarranted conversion of local evaluation into interpretive standing: a result genuinely passes a local check within a specific evaluation scope, but is treated by subsequent papers, platforms, registries, or dissemination vehicles as sufficient warrant for a stronger claim of interpretation or scholarly achievement, without the addition of new materials or bridging arguments commensurate with the stronger claim, and when the original material boundaries, uses, and conditions for downgrading or withdrawal no longer actually constrain the subsequent, stronger claim.

``Substitution'' here does not mean that the original evaluation score carried no information, nor that the subsequent, stronger claim is necessarily wrong. What it describes is a change in the relation of reasons: a result that could only support ``a local pass under these materials, this question, and this set of standards'' is rewritten as ``therefore the interpretation has been formed,'' ``the system possesses a certain research capability,'' or ``the result has been established.'' To avoid conflating ordinary extrapolation, inadequate measurement, and exaggeration in dissemination, five conditions are jointly necessary.

First, the local pass in the original evaluation must be genuine. If the citations are fabricated, if factual verification failed, or if the report never satisfied the original criteria, the first diagnosis should be error, hallucination, or evaluation failure; not every failure can be called standing substitution. The distinctive pressure of standing substitution lies precisely in the fact that the original evaluation result can be genuine.

Second, the subsequent level must advance a stronger claim. Rewriting ``passed within this task'' as ``this interpretation is provisionally acceptable'' may not yet cross a level; rewriting it as ``an open interpretation has been formed,'' ``it represents a historical object,'' or ``it demonstrates that the system possesses general research capability'' significantly raises the strength of the subsequent claim. Third, the subsequent, stronger claim must also take the original evaluation pass as its key warrant. If the subsequent claim relies entirely on new historical materials, independent review, and the researcher's own bridging argument, so that the original evaluation score is merely background information, there is no chain of reasons that treats a local pass as the whole being established.

Fourth, the subsequent level has not added commensurate new materials or bridging arguments. Stronger claims require stronger support --- a principle already clearly expressed in Kane's validity argument. New primary materials, version comparison, contextual analysis, competing interpretations, the handling of counterexamples, and causal bridging can all bear this added burden; without them, the original evaluation pass cannot be upgraded into the recognition of an open interpretation. Fifth, the original boundaries, uses, and conditions for downgrading or withdrawal must no longer actually constrain the claim at the subsequent level. If the dissemination materials still explicitly state that the evaluation addressed only a certain task, a certain range of knowledge, and a certain use, and allow new evidence to downgrade the conclusion, standing substitution cannot be diagnosed merely from an enlarged name. ``Actual constraint'' here is not a matter of whether the qualifications still linger in a footnote or appendix; it depends on whether they genuinely change the title, abstract, capability naming, scope of application, dissemination wording, or use decisions at the subsequent level. A counterfactual question can serve as a test: if the material boundaries, uses, and downgrading conditions of the original evaluation were carried into the subsequent level unchanged, would the subsequent, stronger claim have to shrink? If the answer is yes, while the subsequent level still maintains the broader naming of standing, then the original qualifications, though perhaps still verbally present, have lost their actual constraining force.

These five conditions distinguish standing substitution from the general point that ``a proxy is not the target.'' Sharma's summary of Proxy-Based Evaluation has already made clear that metrics, benchmarks, transparency documentation, and auditing procedures are proxy mechanisms of oversight; that the proxies are satisfied still does not prove that the deployment goals have been achieved. This paper does not rediscover this point, but traces how proxy results acquire recognition among academic vehicles. Nor does standing substitution require that any agent deliberately game the metrics. The tradition of metric reactivity represented by Goodhart (1975) and Campbell (1979) mainly warns that the objects of evaluation respond to metrics and institutions; standing substitution, by contrast, can be completed step by step across multiple vehicles without deliberate catering, without score fabrication, and even without any single actor making the complete erroneous inference.

A further boundary must be drawn between the validation of an output and the recognition of a producer or achievement. This paper does not treat those as the same process. An output may pass a specified validation loop without its producer acquiring durable credibility, while an interpretation may receive provisional standing without becoming part of a producer's long-term reputation. The two chains should therefore not be conflated.

\subsection{Two Counterexamples and One Case in Which Standing Substitution Does Not Occur}\label{two-counterexamples-and-one-case-in-which-standing-substitution-does-not-occur}

The first counterexample is a long-form historical report: all its citations are genuine, the pages do exist, and factual verification, source attribution, coverage, and the report's internal logic all pass the established checks. But the report's key causal judgment merely repeats a piece of secondary scholarship, without reworking the counterevidence in the primary materials and without explaining why the secondary author's generalization applies to the present problem. The local pass in the original evaluation is genuine; if the subsequent discourse names it ``a historical interpretation that has been formed,'' it simultaneously lacks new materials and bridging arguments commensurate with the stronger claim, and if the report no longer preserves the boundaries of ``secondary generalization, limited corpus, pending review,'' it constitutes standing substitution. This example shows that standing substitution cannot be exhausted by hallucination or citation error: the citations may all be genuine while the relation of reasons remains insufficient.

The second counterexample is a digital-humanities analysis: the statistics, the co-occurrence network, and the model's numerical values can all be rechecked against public code and data, and the numerical changes remain stable across different runs. A subsequent paper, however, directly names a certain numerical relation a ``conceptual-history event,'' and from it infers a social mechanism of an intellectual tradition. Without sense analysis, controls for chronology and genre, alternative orderings, contrary materials, and a bridging argument from numerical relations to historical practice, the recheckability of the numbers supports only a computational result or a candidate association; it cannot support the stronger conceptual-historical and causal claims. What is genuinely missing here is not more decimal places, but the materials and arguments needed to change the level of interpretation.

It is also possible that no treating of a local pass as the whole being established occurs. Suppose a research system always marks its generated results as ``candidate leads,'' with each result accompanied by the scope of materials, versions, the evaluative use, and an account of expert involvement; and the dissemination documents explicitly state that the report passed only within a specific task and that competing interpretations and new materials can downgrade or withdraw the conclusion. The researchers then add independent primary materials, version comparison, contextual analysis, and the handling of counterexamples, and in the paper they set out, step by step, the bridging process from candidate relations to a historical interpretation. In this case, standing may indeed be legitimately elevated from a local result to a stronger interpretation, but the elevation comes from the added evidence and bridging arguments, not from the original evaluation pass being smuggled in as sufficient warrant; it therefore does not constitute treating a local pass as the whole being established.

\subsection{What Happens After Results Enter Other Settings}\label{what-happens-after-results-enter-other-settings}

The framework of judgment here is a diagnostic framework, not an empirical conclusion already completed for all projects. It can generate three propositions for subsequent research. First, when an evaluation result moves from a system report into papers, platforms, and dissemination materials, the material boundaries, uses, and failure conditions may gradually diminish; this requires paragraph-by-paragraph comparison of versioned documents and cannot be inferred from names alone. Second, the dissemination of results may prefer capability names with broader extensions that are more easily recognized by the public, making local task capabilities more readily understood as general research capability; this is a testable hypothesis about choices in dissemination, not a fact about all projects. Third, the more a project lacks records of boundary migration, the more likely its local evaluations are to be used by later papers, platforms, or dissemination materials for stronger claims of standing; this requires cross-project comparison of evaluation records, dissemination texts, and audience use, and cannot be proven by this paper's conceptual analysis alone.

These propositions do not prearrange open-ended and closed tasks into risk grades. Closed questions can measure bounded capabilities accurately, and open-ended tasks can use expert criteria to yield meaningful local information; what really needs to be tracked is whether a pass within the evaluation scope is endowed with a new claim when it leaves the original task, and whether the original constraints move along with it. Espeland and Stevens's (2008) discussions of quantification and commensuration, Power's (1997) warning about the audit society, and Muller's (2018) analysis of metric fixation can provide a sociological vocabulary for this institutional background, but none of them can replace the examination of concrete vehicles, versions, and chains of reasons.

\subsection{Conclusion: A Local Pass Is Not an Established Interpretation}\label{conclusion-a-local-pass-is-not-an-established-interpretation}

The scope and basis of evaluation make comparison possible, and they also provisionally fix materials, tasks, scoring, and failure conditions as objects the community can check. Its value lies not in compressing open interpretation into a score table that never changes, but in making the scope and reasons of a local judgment statable. Kane (2013) and Messick (1989) have shown that claims about interpretation and use must accept evidential scrutiny commensurate with their strength; research on complex evaluation has further separated facts, sources, coverage, analysis, and report logic into diagnosable relations. Sharma's (2026) discussion of proxy-based evaluation reminds us that satisfying a proxy does not equal achieving the target; the same distinction matters here because an output passing a bounded evaluation does not by itself confer higher status on the producer or the result.

The increment of this paper is therefore limited and concrete: it traces how a local evaluation is converted into interpretive standing among papers, platforms, registries, and dissemination. Only when the original evaluation is genuine and valid, the subsequent claim is stronger, the original evaluation is used as the key sufficient warrant, no commensurate new evidence or bridging argument has been added, and the original boundaries and downgrading conditions no longer constrain the subsequent level is there reason to diagnose standing substitution. If the scope and basis of evaluation can be re-examined by the community, and if new materials, counterexamples, and bridging arguments do change the claim, then an elevation of standing can hold; if these conditions appear only in words without being able to change the conclusion, one must still return to the responsibility layer and ask whether the judgment is genuinely questionable, revisable, and withdrawable.

\section{Who Explains and Answers for the Judgment --- Interpretive Credit and Responsibility for Judgment}\label{who-explains-and-answers-for-the-judgment-interpretive-credit-and-responsibility-for-judgment}

\subsection{What Remains After an Interpretation Has Been Recognized}\label{what-remains-after-an-interpretation-has-been-recognized}

The standing substitution discussed in Section 3 concerns how a local evaluation, in cross-vehicle circulation, comes to be endowed with a stronger interpretive status; it does not automatically follow that a structure of responsibility is missing. A researcher may, after a bounded evaluation, add materials, engage counterexamples, and make the bridging reasons public, so that the original local pass becomes one link in a stronger judgment. Conversely, even if an interpretation has been provisionally recognized by journals, platforms, or peers as citable, teachable, or usable as a basis for further work, the interpretive credit it carries may no longer be connected to a person or structure capable of answering for the overall judgment. What this paper means by ``interpretive credit'' is only the epistemic and institutional credit that a text or a signed achievement acquires because it is regarded as ``an interpretation that has been formed''; it is not a synonym for interpretive truth, nor is it identical with the long-term prestige acquired by an author.

The question here is therefore neither whether a generative system can become a moral agent, nor a post hoc search for an individual who can bear blame. A judgment in humanistic scholarship can acquire provisional authority usually also because it is placed within a practical relation to which one can return: someone can indicate the scope and versions of the materials, explain what role a given piece of material actually plays in the argument, and account for the intermediate reasons leading from materials to concepts and from concepts to historical generalization; others can accordingly advance contrary materials, different contexts, or competing interpretations. If this relation is deleted, the finished product may remain coherent and the genuine citations may still exist, but what the community receives is a judgment already completed, not a judgment whose formation, reasons, and limits can still be queried.

A distinction should be drawn here between ``there were in fact people behind the conclusion'' and ``the relations of responsibility have been made public.'' A paper produced by multiple collaborators, through model generation and editorial processing, may of course have many actual participants; the question is whether readers can tell which agents or institutional links are responsible for which judgments, whether they can demand reasons from them, and whether they can know under what counterexamples, version problems, or insufficiencies of evidence the conclusion will be downgraded, rewritten, or withdrawn. What is to be identified here is not some hidden psychological fact, but whether this relation has, in the public record, an identifiable form sufficient to support interpretive credit.

\subsection{Footnotes, Context, and the Constraint of Materials}\label{footnotes-context-and-the-constraint-of-materials}

Grafton's account of the history of the footnote provides a relatively concrete point of entry. The footnote has never been merely a typographic device that decorates existing knowledge into professional knowledge: it connects the author's assertions to materials that can be sought out, checked, and differently interpreted, and it places the author under the continuing scrutiny of peers and readers. The early historiographical controversies Grafton cites make this especially clear: listing ``proofs'' does not oblige readers to agree with the author; precisely because the materials are indicated, readers may treat the same materials in different ways and demand that the author revise. The value of the footnote therefore lies not in guaranteeing truth, proving how much labor the author actually expended, or replacing complete argument, but in making the assertion a challengeable commitment: it must accept questions about whether the sources genuinely support it, whether omissions change the conclusion, and whether citations have been selectively used.

This also shows that the mere existence of footnotes cannot satisfy the conditions of responsibility. Footnotes can be piled up, can display only the range of one's reading, and can miswrite secondary generalizations as primary evidence; a complete citation interface may even strengthen the appearance that an interpretation has been completed. Only when readers can follow the footnotes back to the materials, and when this return can change the acceptability of the argument, do footnotes participate in responsibility for judgment. This paper does not understand such a return as the recovery of an absolutely infallible evidential foundation; it requires only that the judge cannot, after the citations have been challenged, continue to maintain the conclusion as an established fact unaffected by the challenge.

Skinner's critique of method in intellectual history gives this point another qualification. He opposes treating the text as a self-sufficient object, and he also opposes treating the macro-social context as a background that can directly determine the text's meaning; to understand an utterance, one must examine the speech acts conventionally performable on that occasion, and make the reconstruction of vocabulary, questions, audiences, and modes of expression a constraint among competing interpretations. This paper draws no ontological conclusions about artificial intelligence from this; it borrows the account only to show that humanistic interpretation does not confront a textual surface that can be filled in at will. The dating of materials, the availability of words at the time, the historical position of the topic, and neglected alternative formulations can all deprive an apparently smooth generalization of its conditions of holding. Context here is not supplementary background material; it is the constraint that materials impose on judgment, and when new contextual evidence appears, the original judgment should be open to change.

\subsection{Judgment Is Not Shown by Unverifiable Inner Processes}\label{judgment-is-not-shown-by-unverifiable-inner-processes}

If responsibility is understood merely as whether the signatory can answer questions after the fact, the temporality of the formation of judgment is still missed. At this point the paper makes limited use of Gadamer: understanding is not a one-directional act in which an already completed subject imposes meaning on an object; the process of understanding also changes the questioner's own ways of asking and anticipating. This dimension of formation does not require researchers to disclose all their private reflection, still less does it equate visible process directly with genuine understanding; it only suggests that the person or structure capable of answering for a judgment should allow materials, discussion, and counterexamples to change the conclusion it sets out to defend.

The internal standards of practice emphasized by MacIntyre supply the public face of this point. Competence is not an endowment possessed once and for all, detached from practice; it is formed gradually within the standards of a shared activity, the experience of failure, and the testing of a community. This paper does not turn this view of practice into a threshold of ``only those unassisted by AI can form judgment''; human--machine collaboration can likewise form a structure of responsibility within a clear division of labor, community testing, and withdrawable conclusions. What it rejects is only another over-hasty inference: that so long as a text is formally proficient, or so long as someone signs the cover, one can dispense with an account of how the judgment acquired its standing through materials, failures, and revisions.

Gadamer and MacIntyre perform only one delimiting task in this paper: the former shows that understanding in formation is reshaped by its object and by the process, and the latter shows that the capacity for judgment must undergo failure within testable practical standards. Neither can adjudicate whether machines understand or whether they can be responsible, nor can either prove that any particular user has not understood. What this paper needs is not a chain of reasoning from philosophical anthropology to a technological prohibition, but a narrower institutional judgment: if interpretive credit is to continue to be recognized, the fact that a judgment, in the process of its formation, remains open to being changed by materials, discussion, and counterexamples cannot wholly disappear from public relations of responsibility.

\subsection{Why Errors, Hesitations, and Drafts Matter}\label{why-errors-hesitations-and-drafts-matter}

Adhya's (2026) observation of the classroom is directly, though limitedly, instructive here. She notes that when learning is compressed into a submittable product, fluency, orderliness, and finish may conceal the fact that reading, discussion, and revision have not occurred; and the errors, hesitations, drafts, annotations, and discussions of in situ writing are worth including in evaluation not because imperfection is inherently more authentic, but because they sometimes preserve traces of a judgment still in formation, still open to being changed by others. This observation cannot be directly transferred as an empirical conclusion about the research community: drafts can be forged, hesitation can be mere irrelevant delay, and a refined final draft may equally well rest upon rigorous revision.

This paper therefore does not take ``leaving traces of process'' as a sufficient condition for an interpretation to be established. Process materials are relevant to responsibility only when they can show how materials entered the judgment, how counterexamples were handled, which intermediate judgments were abandoned, and what conditions would trigger downgrading or rewriting. What Adhya (2026) resists is the complete replacement of the learning process by the product; what this paper further asks is whether, when the product leaves the classroom and enters papers, databases, platforms, or public circulation, there is still some public structure by which interpretive credit can be reconnected to the process in which the judgment was formed. The two are connected, but the pedagogical observation of process is not thereby expanded into suspicion of all academic texts.

\subsection{Who Explains the Judgment, and Who Bears the Consequences}\label{who-explains-the-judgment-and-who-bears-the-consequences}

This paper uses ``responsibility for judgment'' as the general term, but does not conflate it with \emph{answerability} or \emph{accountability}. The former means that the bearer can be required to state the reasons for a judgment and to respond to materials, inferences, and counterexamples; the latter means that such accounting, correction, downgrading, withdrawal, and the bearing of consequences have been embedded in some executable institutional relation. Responsibility for judgment is the wider practical relation: who advances and maintains a judgment, on what reasons it is maintained, which counterexamples can change it, and who undertakes the corresponding actions when the judgment needs revision.\footnote{\emph{Answerability} is used here in the sense of being answerable for one's judgment---being able to give reasons and to respond to objections---not in the sense of a system's capacity to respond to users.}

On this basis, the third level can be defined as follows: an output, even after it has acquired recognition, may have no public, identifiable structure of responsibility through which its overall judgment can be explained, questioned, revised, or withdrawn. In other words, \emph{An output may acquire interpretive standing even when no publicly identifiable structure of responsibility exists through which the judgment as a whole can be justified, challenged, revised, or withdrawn.} The divergence here occurs between interpretive credit and responsibility for judgment, not between two ontologies called ``human'' and ``machine.''

The five conditions of the responsibility layer jointly specify a re-examinable relation of responsibility. Different disciplines may adopt different procedures of material preservation, authorial division of labor, anonymous review, and errata, and an exploratory judgment may be provisionally advanced before the evidence has reached its strongest state.\footnote{The five conditions are not a responsibility scale that can be scored once before publication, and they presuppose neither a single archival format nor equal responsibility for every participant.} The problem is this: a conclusion has acquired stable credit through publication, platform, or institutional endorsement, yet readers cannot discern how responsibility is allocated, cannot demand reasons from the bearer, and cannot make new materials actually change that credit. The responsibility conditions do not freeze an interpretation into a qualified product; they preserve the relations in which an interpretation, after acquiring standing, can still be re-examined.

The core of the responsibility relation consists, first, of three conditions: the judgment must be attributable, the reasons must be statable, and counterexamples must be able to change the judgment. First, the judgment must be attributable. A signature, a project name, or a model name is by itself insufficient to complete attribution; readers should be able to tell whether the key judgments were advanced and maintained by an individual, a team, journal editors, a research institution, or a human--machine collaborative structure with an explicit division of labor. Second, the reasons must be statable. The bearer need not hand over every private note, but should be able to state the scope of materials, versions, the roles of evidence, the bridging inferences, and the key alternative paths not adopted. Third, counterexamples must be able to change the judgment. If feedback can only be collected without being able to affect the argument, the evaluation, or the conclusion, so-called responsiveness is merely a form.

Beyond these three core conditions, two executive conditions are needed for responsibility to take effect institutionally. Fourth, the conclusion must be capable of being downgraded or withdrawn. This does not require researchers to suspend every provisional judgment indefinitely; it requires that versions, qualifications, and withdrawal conditions remain visible while the conclusion still has practical effect, so that ``candidate leads,'' ``bounded generalizations,'' and ``interpretations already defeated by counterevidence'' do not continue to circulate with the same standing. Fifth, there must be a person or an explicit structure that bears the consequences. The consequences here include scholarly errata, reputation, and re-evaluation, and also the answerable effects produced when a judgment is used in curricula, research decisions, or public discourse; this does not presuppose that all consequences should be borne by a single individual, but it excludes the situation in which responsibility is diluted without limit among system, platform, authors, editors, and institutions.

What the five conditions jointly specify is the practical form of responsibility, not an individualist model of ownership. An individual author can satisfy it; a research team can satisfy it through a clear division of labor; journals, laboratories, or archival institutions can bear some of its conditions; in human--machine collaboration, the system can participate in retrieval, generation, comparison, and revision, while it must still be made clear who is responsible for explaining the overall reasons, who has the authority to revise or withdraw, and what procedure enables counterexamples to change the judgment. The bearers of responsibility may be plural, but the structure of responsibility must not thereby become unidentifiable. Conversely, if a team clearly marks, in the public record, the role of the model, the boundaries of the materials, the points of human judgment, and the withdrawal procedure, the use of a generative system by itself does not constitute a deficiency of responsibility.

The re-examination of code usually occurs not at the first run, but in maintenance, requirements changes, dependency upgrades, vulnerability disclosures, production incidents, and system migrations. What then needs to be re-explained is not only whether a function returns the expected result, but also why this architecture was chosen in the first place, which evidence would force it to change, and who has the authority to modify, suspend, roll back, and withdraw. A legacy system that runs but that no one can explain and no one can safely modify displays precisely the code variant of the responsibility layer: the artifact has acquired institutional standing, while the overall judgment lacks a public structure of responsibility through which it can be explained, questioned, revised, and withdrawn. A successful run has not eliminated the problem of responsibility; it has merely postponed the moment of re-examination to the stages of maintenance and incident response.

\subsection{The Distance from Neighboring Responsibility Problems}\label{the-distance-from-neighboring-responsibility-problems}

This definition differs, first, from the question posed by the responsibility-gap literature. The classic formulation of Matthias (2004) and the extended analysis of Santoni de Sio and Mecacci (2021) mainly address how fault, accountability, and active responsibility should be ascribed after the behavior of learning automata and its adverse consequences have occurred. This paper of course borrows their wariness toward a single answer to responsibility ascription, but it does not discuss the moral and legal imputation of accidents, harms, or automated behavior. What this paper asks about is another moment in time: after an interpretation has acquired standing in academic or public circulation, who is responsible for its structure of reasons as a whole. The former can occur in automation scenarios in which no interpretation is recognized at all; the latter can occur in academic writing in which there is no imputable harmful event. The two are structurally similar but cannot replace each other.

A further self-limitation concerns agency and authorship. This paper does not measure genuine human agency, construct an agency score, or reduce distributed scholarly work to a single authorial signature. Its object remains narrower: whether an interpretation in circulation has a public structure of reasons, revision, and withdrawal. Authorship and signature therefore show neither that responsibility is identical with single authorship nor that a signature alone completes the responsibility of judgment.

These two sets of distinctions at the same time delimit this paper's diagnosis. The responsibility layer does not consist in discovering that no human being participated in the interpretation, nor does it infer from a finished text that some author in fact did not read, reflect, or hesitate. It requires only that, on occasions where the community grants credit to an interpretation, the public record can answer five practical questions: who, or what structure, advanced the judgment; on what reasons it is maintained; which counterexamples can change it; when it can be downgraded or withdrawn; and who bears the ensuing consequences. If these questions cannot be answered at the public level, there is reason to say that interpretive credit and responsibility for judgment have come apart; if they can be answered, the interpretation may still be wrong, but it should not be diagnosed as a deficiency of responsibility merely because generative tools were used.

\subsection{Conclusion: Responsibility Is Not a Label Attached to a Finished Product}\label{conclusion-responsibility-is-not-a-label-attached-to-a-finished-product}

The reason interpretive standing cannot be automatically granted by textual finish, genuine citations, or local evaluation scores is not only that these indicators may still miss errors. The deeper difficulty is that they can allow a judgment to circulate before any public structure of bearing has been established: what readers see is an interpretation already completed and thus seemingly already answered for by someone, while what is actually visible may be only the product, the signature, or the evaluation label. The footnote's commitment as revealed by Grafton (1997) and the contextual constraint demanded by Skinner (1969) remind us that an interpretation must face the counter-pressure of materials and readers; Gadamer (2004), MacIntyre (2007), and Adhya (2026) respectively show, from the formation of understanding, practical standards, and traces of process, that a judgment is not a static product that can be wholly separated from the possibility of being changed.

A structure of responsibility can exist while the judgment remains wrong, and footnotes, version records, and community procedures may be executed only formally. What is required here is that interpretive credit be connected to the structure of bearing described in this section; where that structure is missing, it should not be treated as an already stabilized interpretive standing. The next section will explain why humanistic scholarship makes this requirement especially visible: what it needs is not the indefinite postponement of conclusions, but the capacity to keep the closure of an interpretation open to re-examination by materials, concepts, and counterexamples.

\section{How Humanistic Scholarship Tests the Conditions Under Which an Interpretation Is Established --- The Formation of Judgment and the Revisability of Conclusions}\label{how-humanistic-scholarship-tests-the-conditions-under-which-an-interpretation-is-established-the-formation-of-judgment-and-the-revisability-of-conclusions}

\subsection{Humanistic Scholarship Does Not Refuse Evaluation}\label{humanistic-scholarship-does-not-refuse-evaluation}

The distinctions of the preceding three sections do not require imagining humanistic scholarship as an activity naturally superior to technical evaluation. Historical interpretation, textual explication, and conceptual-history research likewise require a scope of materials, a basis in versions, norms of citation, peer review, and responses to counterexamples; a beautifully written article that cannot explain where its materials come from or what role its citations play cannot acquire recognition on the strength of its style alone. To treat humanistic scholarship as a testing ground does not mean that it rejects standards; it means that it more often exposes a relation that standardized evaluation cannot settle in one pass: materials may change the question, new contexts may change the boundaries of concepts, and competing interpretations may force researchers to decide anew what counts as relevant evidence.

In laboratory-style closed tasks, the question, the materials, and the scoring criteria can usually be specified first, and only then is the result judged for satisfaction. Humanistic research of course makes similar provisional specifications, but these specifications themselves often become objects of controversy. A difference between versions may dissolve the original ``same text''; a neglected usage of a word may invalidate an apparently clear conceptual distinction; a new archival document may not add one more fact to an old interpretation, but change what the researcher takes to be the phenomenon in need of explanation. The openness here is not the absence of boundaries; it is that the boundaries can still be rearranged in contact with the object.

\subsection{Why an Interpretation Should Not Be Settled Prematurely}\label{why-an-interpretation-should-not-be-settled-prematurely}

This paper calls this state of remaining rearrangeable ``delayed closure,'' that is, not settling an interpretation prematurely: what it delays is not writing, but the moment at which an interpretation is made irrevocable.\footnote{``Delayed closure'' is a working name coined for this paper; it does not claim to be a stable term in hermeneutics or humanistic scholarship, and it bears no genealogical relation to premature closure in the cognitive sense or to closure principles in epistemology. Only the sense of ``standing remains re-examinable'' is intended.} A conclusion may hold provisionally on the current evidence, may be cited, or may be used to pose the next question, but its material boundaries, competing interpretations, unhandled counterexamples, and withdrawal conditions must remain visible, and new evidence must genuinely be able to change it.

Delayed closure is therefore a combination of practices, not an index of time. Version comparison lets researchers know whether the text before them is the same; the return through footnotes re-exposes citations to sources and to readers' different handling; conceptual discrimination keeps a familiar word from sliding silently across epochs; contextual rechecking keeps the present question from arbitrarily overrunning the historical object's own questions; and the handling of counterexamples lets negative materials actually change the scope of the claim rather than merely appearing in a ``limitations'' section. The capacity to downgrade, suspend, and withdraw is the condition under which these practices have institutional consequences. A ``reservation of judgment'' without consequences may still be only part of the finished appearance of the output.

\subsection{Leaving Room for Revision So That Judgment Can Form}\label{leaving-room-for-revision-so-that-judgment-can-form}

The formation of judgment is not a matter of there first being a complete subject who then projects meaning onto passive materials. In working through materials, a researcher may find that the original question does not hold, or that a different set of concepts must be adopted; the community's objections may deprive evidence that once seemed sufficient of its force. It is precisely in this process --- affected by the object and by others, and continually changed --- that a judgment gradually takes on a form that can be explained and revised. The ``formation'' spoken of here does not appeal to unverifiable inner processes; it refers to a relation that can be partially tracked in the public record: which materials entered the judgment, which materials were excluded, which counterexample forced the conclusion to be downgraded, and which bridging reasons remain provisional.

This also explains why technical processing can precede judgment but cannot replace it. OCR, chunking, retrieval, and ranking can make some materials appear earlier and make others harder to bring into view; a generative system can also rapidly organize local associations into a complete narrative. But research judgment is not merely choosing the smoothest line among candidate materials; it must also withstand the moments when materials fail to meet expectations, and allow the question itself to be rewritten. Delayed closure does not refuse technical processing; it requires that technical processing leave entry points for re-ranking, adding materials, and changing the claim. If a system preserves only the final coherent narrative and cannot say what materials would stop it, turn it, or withdraw it, then the formation of judgment has been concealed by the product.

\subsection{Why Generative Systems Make the Problem More Visible}\label{why-generative-systems-make-the-problem-more-visible}

Generative systems can rapidly produce texts that pass within existing evaluation systems: they can organize facts, citations, paragraph structure, and a cautious tone into a single report, so that a stage judgment quickly acquires the appearance of completion. The normative risk here is not the empirical assertion that ``machines necessarily produce middling texts,'' but that institutions may prefer a form of result that is easy to compare, easy to disseminate, and easy to archive. When a qualified finished appearance is repeatedly treated as a stable proxy for research capability, researchers may reduce those practices that temporarily lower clarity but allow the question to be rewritten. What is ultimately crowded out is not just some erroneous step, but the space for posing a different question, keeping undecided materials in view, and admitting the failure of an interpretation.

The value of humanistic scholarship should therefore not be reduced to preserving handwritten drafts or maintaining a purely human style. A machine-assisted research process can entirely form responsible judgment, so long as it lets materials interrupt an existing narrative, lets counterexamples change evaluations, and keeps the judgment nodes and withdrawal authorities of different participants visible. Conversely, a hand-written text full of hesitations may hide all its qualifications. What leaving room for revision protects is the possibility that the exploratory process can be readjusted, not the purity of some medium or authorial identity.

The significance of this test lies precisely here: it brings back into view the temporality of formation that can be provisionally set aside in most evaluation environments. Scoring often requires that a judgment present itself at some deadline as a comparable result, and publication and dissemination require that the result bear a name that can be briefly restated; humanistic research, by contrast, often must maintain a working tension between completion and incompletion. When materials are insufficient, the most responsible result may be to shrink the claim; when a concept remains ambiguous, the most valuable advance may be to change the question rather than to complete the conclusion; when counterexamples have not been handled, suspending judgment is not an evaluative failure but a way of avoiding writing a candidate interpretation prematurely into public fact. If institutions reward only answers that have been closed, these actions will be recorded as insufficient efficiency, and interpretive standing will favor the texts that most easily present themselves as finished.

This does not mean that all uncertainty is worth preserving. Hesitation may indicate insufficient evidence, but it may also be an excuse for not having completed the necessary work; openness can likewise be used to evade responsibility for materials and reasons. Delayed closure must therefore attach to concrete, checkable actions: indicate which kind of material is still missing, state which counterexample remains unresolved, mark what evidence would downgrade the conclusion, and actually revise the text when new evidence arrives. Only then is delay not the indefinite postponement of standing, but a way of keeping the acquisition, maintenance, and withdrawal of standing explicable.

Code --- and in particular AI-assisted coding practices that take run results as their main feedback, colloquially ``vibecoding'' --- provides a reverse test. Compilation, execution, and testing can quickly close local functional questions, but they cannot in the same way close questions of requirements, architecture, and responsibility; from passing tests to a system ready for production, one must still pass through specification interpretation, untested conditions, security review, version records, and maintenance authority. The point of conversion here is therefore very clear: the problem of appearance has not disappeared; it has migrated from whether the code works to what overall standing can be inferred from this working result.

The difference between humanistic scholarship and code is therefore not a difference between substance and appearance. Humanistic interpretation lacks a single public runtime that can quickly and repeatedly exhaust its meaning-claims, and the cycles in which materials, versions, and counterexamples change judgments are usually longer; software engineering, through version control, code review, versioned testing, rollback, and maintainer institutions, embeds part of the requirement of re-examination in engineering practice. The latter set of institutions is not a sufficient guarantee either: only when they can actually change evaluations, block deployment, initiate revision, or locate consequences do they perform the function of delayed closure. This comparison is not a scale for ranking disciplines; it shows that the same problem of standing appears at different positions under different verification conditions.

The athlete example from the introduction can be seen more clearly within this comparison. A goal or a beautiful move can be confirmed on the spot, but cannot by itself prove a stable ability; as far as on-the-spot confirmation is concerned, the run result of code is similar. But there is a deeper difference between the two: the athlete's conditions of formation are enforced by the body --- training, failure, and correction cannot be generated or forged in the same symbolic medium as the performance --- and the finished appearance and the process of formation are in principle inseparable. The artifacts, tests, comments, and commit records of code, by contrast, all reside in a medium that language models can natively generate, and the record of formation itself can become part of the finished appearance; the runtime environment only partially restores the enforcing function of the body, and it covers only the functional layer. Humanistic interpretation lacks even this layer of substitution; its resistance to divergence can come only from materials, counterexamples, community controversy, and the concrete practices discussed in this paper. Delayed closure thereby acquires its exact position: in the absence of body and runtime, humanistic scholarship, if it is to keep interpretive standing re-examinable, can only maintain this substitute structure on its own.

\subsection{From Three Questions to a Repeatable Process of Testing}\label{from-three-questions-to-a-repeatable-process-of-testing}

Placing delayed closure back into this paper's three-layer framework makes its institutional function clearer. The formation layer requires that the record allow return; the evaluation layer requires that a local pass remain constrained by the contract, the use, and the permitted inferences; the responsibility layer requires that a judgment that has acquired standing remain connected to an identifiable structure of bearing. Delayed closure is not a fourth independent threshold, but the common condition that keeps the relations among these three layers reversible: records of formation can prompt re-evaluation, evaluative controversies can push a claim to be rewritten, and responsibility procedures can prompt a conclusion to be downgraded.

This cycle also shows why transparency by itself is not enough. Publishing more logs, if the logs can only prove that the system ran but cannot show how materials changed the judgment, cannot repair the formation layer; publishing scores, if the scores are still treated as general capability after leaving the original task, cannot repair the evaluation layer; listing a responsible person, if that person has no authority to revise or withdraw the judgment, cannot repair the responsibility layer. What delayed closure requires is relational revisability, not an increase in the quantity of information. It can be realized through version records, registries of counterexamples, revisions of the scope and basis of evaluation, and explicit withdrawal procedures, but these forms have interpretive significance only when they can genuinely change subsequent standing.

\subsection{A Further Distinction Concerning Distributed Agency}\label{a-further-distinction-concerning-distributed-agency}

A distinction must also be kept between this paper's institutional question and broader theories of distributed agency in AI-mediated writing. This paper does not attempt to measure ``genuine human agency,'' nor does it construct an agency score capable of comparing different writing subjects. What this paper means by delayed closure is a concrete set of practices: whether the community can trace the conditions of formation, whether the evaluation contract can be re-examined in light of new materials and counterexamples, and whether the structure of responsibility can demand explanation, correction, and withdrawal.

This distance is also a self-limitation. If future research shows that a distributed system can preserve adequate records of formation, allow evaluation criteria to change, and bear the consequences of judgment, this paper has no reason to deny its interpretive standing merely because machine participation is involved; if a human author's text has no structure of reasons open to re-examination, the signature alone cannot automatically satisfy the three layers of conditions. Delayed closure provides only a way of testing public arrangements; it does not provide a final answer about consciousness, personhood, or human uniqueness.

\subsection{Conclusion: Keeping Conclusions Open to Revision}\label{conclusion-keeping-conclusions-open-to-revision}

Humanistic scholarship is an especially telling test of this problem not because it enjoys a privilege of exemption from evaluation, but because it pushes to the foreground those relations in the formation of judgment that are easily concealed by the finished product: materials resist existing questions, contexts restrict the migration of concepts, counterexamples change conclusions, and researchers change their own understanding in the process. Generative systems make it easier for the finished appearance of the output to circulate first, and therefore make it all the more necessary to preserve the possibility of revision and withdrawal, lest the product be prematurely treated as an unchangeable conclusion.

What this paper means by delayed closure ultimately points to a limited normative requirement: stage conclusions may be advanced, but they cannot continue to circulate at the same strength after losing their material boundaries, evaluative uses, and structures of responsibility. It does not demand infinite slowness, nor does it treat errors, hesitations, drafts, or human signatures as guarantees of truth; what it requires is that when new materials, counterexamples, or community controversies appear, an interpretation still have a real path of re-examination. Section 6 converts this requirement into five more concrete public conditions, while preserving their character as non-sufficient guarantees.

\section{Five Public Requirements --- From Principles to Practices}\label{five-public-requirements-from-principles-to-practices}

\subsection{Making ``Revisability'' a Public Practice}\label{making-revisability-a-public-practice}

The following five requirements are not a uniform technical checklist that all disciplines must adopt; they are minimum requirements commensurate with the risk of the judgment. They can be realized through different forms --- footnotes, version notes, research logs, editorial procedures, or platform markers; the forms may differ, but they must bring key materials, bridging reasons, limiting conditions, and revision authority back within the community's range of inspection.

If the delayed closure discussed in Section 5 remains at the level of an attitudinal expression that ``research should stay open,'' it is still insufficient to change how an interpretation circulates in the community. A system may declare in its documentation that its results are provisional, and a paper may list limitations at its end, but if readers cannot know where the materials came from, what translations the judgment passed through, or when the conclusion should be downgraded, the professed openness will impose no actual constraint on the interpretive credit already acquired. Re-examination is therefore not an ethical reminder appended to the finished product; it is the preservation, within the result and its subsequent papers, platforms, or dissemination materials, of the re-examinable relations among formation, evaluation, and responsibility.

The minimum objects of the public conditions are the records directly relevant to interpretive standing that the community can return to: material boundaries and versions, the roles of evidence in the argument, the translation between model-generated content and the researcher's judgment, the applicable scope of the evaluation contract, counterexamples and failure states, and the structure of responsibility with authority to revise or withdraw the conclusion.\footnote{``Public'' does not require disclosing all prompts, internal model states, private notes, or unprocessed chains of thought, nor does it turn the research process into continuing surveillance of individuals.} So long as these relations remain traceable and can actually take effect when necessary, the records may take different technical and disciplinary forms.

\subsection{Requirement One: State Which Materials and Which Versions Were Used, and Whether Others Can Access Them}\label{requirement-one-state-which-materials-and-which-versions-were-used-and-whether-others-can-access-them}

An interpretation cannot carry only a string of source names; it must also state which materials actually entered the judgment, in which versions the materials stand, and whether readers, at the time the judgment was made, could return to them under the same or comparable conditions of access. ``Scope'' includes not only the literature, archives, corpora, and data that were included, but also the categories of excluded materials and the reasons for exclusion; ``version'' includes not only file dates, but also revisions, translations, abridgments, scan quality, database updates, and retrieval times --- conditions that change the identity of the evidence; ``conditions of access'' concern permissions, link rot, regional restrictions, changes in retrieval interfaces, and whether the model could actually read the materials. Without these qualifications, the same statement ``cited a certain work'' may represent different evidential relations at different times, in different versions, and under different conditions of availability.

This requirement does not mistake repeatable access for a sufficient foundation of interpretation. A fully open dataset can still be misunderstood, and an archive with complete versions still requires contextual and conceptual judgment. What it prevents is another, more basic divergence: the product retains the names of sources while readers cannot know how the material boundaries affected the conclusion, or cannot tell whether a later version has replaced the original basis. For generative systems, the minimal conditions of retrieval results, context truncation, and model version should also be recorded; if these cannot all be disclosed for reasons of privacy, copyright, or security, one should at least state which inferences lose traceability because of the missing parts, and correspondingly lower the strength of the claim.

\subsection{Requirement Two: Record Facts, System Outputs, and the Researcher's Interpretation Separately}\label{requirement-two-record-facts-system-outputs-and-the-researchers-interpretation-separately}

Re-examination also requires that different evidential levels be kept separate in the record. A textual fact is a statement locatable in the materials; a model result is a candidate output produced by the system under specific prompts, contexts, and parameters; a textual interpretation is an argument about words, structures, and relations of meaning; a historical generalization is an induction about relations across texts, periods, or practices; and a causal mechanism further claims how a relation arises, is maintained, or changes. These may be interrelated within the same study, but they do not acquire the same interpretive standing merely because they are smoothly written together in the final paragraphs.

The function of this distinction is not to prescribe a rigid pipeline, but to make cross-level bridging a checkable object. A model's indication that two words co-occur in a corpus can serve first only as a model result or a candidate association; if the researcher wants to write it as a textual interpretation, they must explain how word senses, genre, and context support this relation; if it is further generalized into a historical mechanism, they must also account for chronology, alternative materials, and the inference from association to mechanism. Each translation may change the scope of the question and the burden of evidence, and should therefore leave, in the version record or an argument map, nodes from which one can return to the original materials and the original judgment. Conversely, labeling a model-generated historical generalization as a ``fact,'' or placing a model inference that the researcher has not rechecked at the same level as a quotation from the source text, lets the finished appearance of the output conceal the conditions of formation.

Layering does not require pre-separating human judgment and machine output into two pure blocks. A researcher may re-retrieve materials on the basis of candidate leads proposed by the model, and a model may complete summaries and comparisons within a conceptual framework specified by humans; what genuinely needs to be marked is which judgment was advanced by which materials and which participant, and which step remains only a bridge awaiting testing. In this way, machine participation by itself is not treated as disqualifying, nor is the coherence of machine prose treated as a completed interpretation.

\subsection{Requirement Three: Failure, Downgrading, Suspension, and Withdrawal Must Actually Work}\label{requirement-three-failure-downgrading-suspension-and-withdrawal-must-actually-work}

If an evaluation regime can produce only ``pass'' or ``generation complete,'' any promise of openness is easily flattened in circulation. Re-examination requires that states such as miss, insufficient evidence, unresolved counterexample, inaccessible materials, task mismatch, and inability to judge be written as formal results, not as noise awaiting cleanup. A system's failure to find materials supporting a claim may mean that retrieval failed, or it may mean that the claim itself needs to be abandoned; neither can be automatically rewritten as ``pending polish.'' Likewise, if a counterexample means that the original conclusion can be retained only within a narrower range, downgrading is the judgment successfully responding to the materials, not a shameful record of declining system performance.

This requirement also concerns the institutional status of suspension and withdrawal. Suspension is not hiding the result; it is explicitly stating that the current evidence is insufficient to maintain the original standing, while preserving what materials could restart the judgment. Withdrawal is not erasing a text once published; it is stopping the old version's interpretive credit from continuing to circulate at its original strength, and letting readers see the reasons for withdrawal, the scope of impact, and the replacement conclusion. Only when these states can trigger version updates, citation notices, database flags, evaluation reruns, or editorial procedures do they possess actual revisability. If ``limitations,'' ``pending verification,'' and ``provisional'' never change the title, abstract, signature, or mode of dissemination, they are merely rhetorical buffers within a finished text.

One must still avoid romanticizing failure here. A miss does not automatically signify depth, and hesitation does not automatically signify responsibility; a team can use openness to evade verification it ought to have done, and can also design withdrawal procedures so cumbersome that they exist in name only. The criterion is whether failure states are connected to concrete evidential gaps, counterexamples, and acts of revision, and whether they can genuinely change interpretive standing --- not whether they look humble.

\subsection{Requirement Four: The Evaluation Contract Must Be Versioned and Changeable in Light of New Materials}\label{requirement-four-the-evaluation-contract-must-be-versioned-and-changeable-in-light-of-new-materials}

The scope and basis of evaluation were defined in Section 3 as the operative constraints of materials, tasks, scoring, permitted inferences, and failure conditions. If a contract has only an initial version, and the applicable scope of a result is no longer recorded once it passes, the local evaluation acquires an unrestricted extension in circulation. Re-examination requires preserving at least the contract's version number, date of formulation, snapshot of materials, scoring dimensions, permitted inferences, and failure handling; when new materials, counterexamples, task purposes, or community criticism change these elements, a revision should be issued, stating within what range the results under the old version remain valid.

Versioning does not replace theoretical judgment with administrative procedure. New materials need not overturn an old contract, and a contract revision need not invalidate all previous results; what needs to be stated is which inference's supporting conditions have changed, which conclusions should be re-evaluated, and which can still be retained as qualified historical records. For open-ended humanistic tasks, a contract revision may especially involve the question itself: a new version or new contextual materials may force researchers to redraw the object, rather than merely adding an item to the existing scoring table. If an evaluation system does not permit modification of the question and the criteria, it cannot record this kind of event that genuinely changes the judgment.

At the same time, the modifiability of the contract cannot become a license to tailor standards to results after the fact. A revision should leave a record of the time, the reasons, and the participants, and state which results before and after the revision are comparable and which are not; when necessary, the original results under the old contract should be preserved, so that the community can ask whether the change of standards was made only to protect existing standing. In this way, the evaluation contract can provide a stable boundary within a single judgment while acknowledging, amid controversy and new evidence, that the boundary itself may need to be redrawn.

\subsection{Requirement Five: State Who Makes the Judgment, Who Explains the Reasons, Who Can Withdraw, and the Status of System-Generated Text}\label{requirement-five-state-who-makes-the-judgment-who-explains-the-reasons-who-can-withdraw-and-the-status-of-system-generated-text}

Finally, the result and its signature must make responsibility for judgment identifiable. At a minimum, it should be clear: who advanced the key judgments; who bears the explanation of reasons and materials; who has the authority to revise, downgrade, or withdraw when counterexamples or version problems appear; and what standing the system-generated text, candidate relations, summaries, or code currently has within the result. The ``who'' here may be an individual, co-authors, an editorial board, a research institution, or a human--machine collaborative structure with an explicit division of labor, but attribution cannot end with only a model name, a project name, or a generic ``team.'' Nor does the structure of responsibility mean pressing all consequences onto the last person to sign: if a platform, institution, or publisher holds substantive authority over review, dissemination, and withdrawal, it too should be recorded within the corresponding scope.

The standing of system-generated content must at least distinguish among candidate leads, facts pending verification, verified citations, arguments rewritten by the researcher, and interpretations accepted by the community. Marking something ``generated by the model'' by itself does not explain what role the content plays in the judgment; likewise, the absence of the model from the signature does not mean that the researcher has assumed the reasons for the sentences the model produced. Only when the researcher can explain how they checked, modified, or rejected the system's content, and has the authority to change the relevant judgments when new evidence appears, is system participation placed within an answerable structure of responsibility. If content is directly incorporated into the result while no agent can explain its bridging inferences or initiate a withdrawal procedure, it may simultaneously reproduce the interpretive appearance of the formation layer and the responsibility gap of the responsibility layer.

A contribution table has effect only when it is connected to the statement of reasons, the version record, and withdrawal authority.\footnote{This paper does not provide cross-disciplinary rules of authorship qualification. A statement of contributions can only mark the division of labor; it cannot by itself complete the attribution of responsibility for judgment.} An anonymous review system can protect critics, and a collective signature can express shared bearing, but both need corresponding editorial, institutional, or team procedures to take up the consequences; otherwise anonymity and collectivity become merely another surface form under which responsibility cannot be located.

\subsection{How the Five Requirements Work Together, and Their Limits}\label{how-the-five-requirements-work-together-and-their-limits}

The requirements above are not a compliance checklist that can be ticked off once before publication. The first provides the path of return to materials; the second makes cross-level translation visible; the third gives negative results actual effect; the fourth lets the evaluation boundary be revised with new evidence; and the fifth connects the judgment and its consequences to an identifiable structure of bearing. The absence of any one may weaken the effect of the others: with complete materials but no layered record, readers still do not know which sentences are model inferences; with a versioned contract but no withdrawal authority, a change of standards still cannot change existing credit; with a responsible person but no accessible materials, responsibility cannot state what its judgment rests on.

The five requirements likewise do not constitute a sufficient condition of interpretive truth. Materials can be misread, bridging arguments can be formally complete yet conceptually fail, and counterexamples may not yet have appeared; a structure of responsibility can honestly acknowledge error, but it cannot thereby turn error into correctness. What this paper can offer is only a narrower institutional proposition: when conditions of formation, evaluation boundaries, and relations of responsibility all have public forms that are traceable, layered, capable of failure, revisable, and attributable, the process by which an interpretation acquires standing is more likely to be discovered, questioned, and changed by the community; when these forms are compressed or disabled in the dissemination of results, the appearance of completion is more easily taken for an established interpretation.

Boundaries of application must also be preserved. Different disciplines' material permissions, anonymity rules, data protection, and scales of collaboration cannot be unified into a single archival format; some historical materials cannot be made public, some model services cannot preserve full run states, and some exploratory projects cannot foresee all counterexamples at the outset. What should be done then is not to pretend the conditions have been satisfied, but to state the gaps and their consequences, and to lower the strength of the claims that can be advanced. If a gap makes a key judgment impossible to trace, question, or withdraw, the most appropriate result may be a candidate lead, internal working material, or suspended dissemination, rather than hiding institutional limits behind a finished report.

\subsection{Conclusion: Keeping Interpretation Open to Re-examination}\label{conclusion-keeping-interpretation-open-to-re-examination}

Re-examining the conditions of interpretation does not mean restoring, outside generative systems, a pure research process unchanged by technical mediation. Materials will still pass through OCR, retrieval, ranking, and model organization, and judgments can be formed by individuals, teams, and human--machine collaboration. The requirement is that these processes must not extract the standing of a subsequent interpretation from its conditions of formation, evaluation, and responsibility. Material scope and conditions of access must be returnable; facts, model results, and interpretations must be distinguishable; misses and withdrawals must genuinely change standing; the evaluation contract must be modifiable by new materials; and the standing of the judgment and of system-generated content must be clearly attributable.

These arrangements increase the likelihood of discovery, contestation, correction, and withdrawal; they do not guarantee that any particular interpretation is true. Nor do they elevate human signatures, handwritten drafts, complete logs, or high evaluation scores into unquestionable credentials. The significance of leaving room for revision lies precisely here: an interpretation may hold provisionally at a given moment, but it must not, merely because its finished appearance has already circulated, lose the real opportunity to be re-examined by materials and by the community. At this point, this paper's three-layer framework forms a bounded cycle: conditions of formation constrain interpretive appearance, the evaluation contract restricts the translation of standing, the structure of responsibility preserves the path of withdrawal, and re-examination keeps these three constraints effective in cross-vehicle circulation.

\section{Conclusion --- An Interpretation Remains Revisable and Withdrawable After It Is Established}\label{conclusion-an-interpretation-remains-revisable-and-withdrawable-after-it-is-established}

This paper began with a seemingly simple puzzle: when a generative system's output has already passed established checks of factuality, citation, coverage, and report logic, why can one still not say directly that a humanistic interpretation has been established? The answer is neither to add another ontological verdict on whether the model truly understands, nor to reduce every problem to hallucination or citation error. Local evaluation can provide genuine and useful information, but it holds only within the evaluation scope constituted by its materials, tasks, scoring, permitted inferences, and failure conditions; once it leaves the original contract and enters papers, platforms, registries, and dissemination, it must face new burdens of standing and responsibility.

This paper delimits this conversion through the relations among three layers. Interpretive appearance, at the formation layer, is the divergence between the finished appearance of the output and the public process of its formation; standing substitution, at the evaluation layer, is the use of a genuine local pass as sufficient warrant for the recognition of a stronger interpretation, without commensurate new evidence and bridging arguments; and the responsibility layer asks whether, after interpretive credit has been established, there is still a public structure that is attributable, able to state reasons, changeable by counterexamples, capable of downgrading or withdrawal, and able to bear consequences. These three layers can also fail separately, and they can reinforce one another in cross-vehicle circulation; they should not be understood as a staircase that every use of artificial intelligence automatically climbs.

The ``establishment'' in the title can therefore only be a withdrawable establishment. An interpretation may provisionally acquire status within specific materials and uses, and may become the working premise of the next step of research, but its standing cannot continue to circulate at the same strength once detached from the process of formation, the evaluation boundary, and the relations of responsibility. The delayed closure of Section 5 is precisely the capacity to keep such standing open to re-examination, and the five requirements of Section 6 propose concrete practices in terms of returning to materials, layering evidence, making failure effective, revising the contract, and attributing responsibility.

Returning to the opening analogy: an athletic performance can ordinarily count as evidence of ability not because the spectators saw the ability itself in a single move, but because there is a traceable connection between the performance and training, failure, rules, and community testing. Generative AI enables the interpretive appearance to acquire a propagational advantage ahead of these connections, and therefore makes it all the more necessary to preserve ``pending verification,'' ``local pass,'' ``candidate interpretation,'' ``suspended judgment,'' and ``withdrawal'' as results that can genuinely change standing. If new materials, counterexamples, or community controversies cannot downgrade, rewrite, or remove a conclusion from circulation, then no amount of fluency, citation, and scoring can show anything more than that an appearance has been maintained.

This does not mean that a human signature, handwritten drafts, complete logs, or machine participation by themselves determine an interpretation's standing. Human--machine collaboration can form responsible judgment, and a human author may be unable to state the reasons they maintain; the key is whether the conditions of formation are traceable, whether the evaluation contract can be re-examined, and whether the judgment is connected to a structure that can explain, revise, withdraw, and bear consequences. This paper's claim accordingly remains limited: interpretive standing cannot be automatically granted by the product, genuine citations, or evaluation scores alone; each time a result moves from one setting into another, one must check anew whether these conditions are still preserved.

\section{References}\label{references}

Adhya, Raahi. ``The Appearance of Interpretation: Teaching Humanities in the Age of AI.'' \emph{Economic and Political Weekly} (EPW Engage) 61, no. 8 (2026). DOI: 10.71279/epw.v61i8.48455.

Bender, Emily M., Timnit Gebru, Angelina McMillan-Major, and Shmargaret Shmitchell. ``On the Dangers of Stochastic Parrots: Can Language Models Be Too Big?'' In \emph{Proceedings of the 2021 ACM Conference on Fairness, Accountability, and Transparency}, 610--623. 2021. DOI: 10.1145/3442188.3445922.

Campbell, Donald T. ``Assessing the Impact of Planned Social Change.'' \emph{Evaluation and Program Planning} 2, no. 1 (1979): 67--90.

Du, Mingxuan, Benfeng Xu, Chiwei Zhu, Xiaorui Wang, and Zhendong Mao. ``DeepResearch Bench: A Comprehensive Benchmark for Deep Research Agents.'' arXiv:2506.11763 (2025).

Espeland, Wendy Nelson, and Mitchell L. Stevens. ``A Sociology of Quantification.'' \emph{European Journal of Sociology} 49, no. 3 (2008): 313--343.

Gadamer, Hans-Georg. \emph{Truth and Method}. 2nd rev. ed.~Translated by Joel Weinsheimer and Donald G. Marshall. New York: Continuum, 2004.

Gao, Tianyu, Howard Yen, Jiatong Yu, and Danqi Chen. ``Enabling Large Language Models to Generate Text with Citations.'' arXiv:2305.14627 (2023).

Goodhart, Charles A. E. ``Problems of Monetary Management: The U.K. Experience.'' In \emph{Papers in Monetary Economics}, vol.~1. Sydney: Reserve Bank of Australia, 1975.

Grafton, Anthony. \emph{The Footnote: A Curious History}. Cambridge, MA: Harvard University Press, 1997.

Liu, Nelson F., Tianyi Zhang, and Percy Liang. ``Evaluating Verifiability in Generative Search Engines.'' In \emph{Findings of the Association for Computational Linguistics: EMNLP 2023}. 2023. 7001--7025. DOI: 10.18653/v1/2023.findings-emnlp.467.

Matthias, Andreas. ``The Responsibility Gap: Ascribing Responsibility for the Actions of Learning Automata.'' \emph{Ethics and Information Technology} 6, no. 3 (2004): 175--183. DOI: 10.1007/s10676-004-3422-1.

Kane, Michael T. ``Validating the Interpretations and Uses of Test Scores.'' \emph{Journal of Educational Measurement} 50, no. 1 (2013): 1--73. DOI: 10.1111/jedm.12000.

Li, Ruizhe, Mingxuan Du, Benfeng Xu, Chiwei Zhu, Xiaorui Wang, and Zhendong Mao. ``DeepResearch Bench II: Diagnosing Deep Research Agents via Rubrics from Expert Report.'' arXiv:2601.08536 (2026).

MacIntyre, Alasdair. \emph{After Virtue: A Study in Moral Theory}. 3rd ed.~Notre Dame, IN: University of Notre Dame Press, 2007.

Messeri, Lisa, and M. J. Crockett. ``Artificial Intelligence and Illusions of Understanding in Scientific Research.'' \emph{Nature} 627, no. 8002 (2024): 49--58. DOI: 10.1038/s41586-024-07146-0.

Messick, Samuel. ``Validity.'' In \emph{Educational Measurement}, 3rd ed., edited by Robert L. Linn, 13--103. New York: Macmillan, 1989.

Muller, Jerry Z. \emph{The Tyranny of Metrics}. Princeton, NJ: Princeton University Press, 2018.

Power, Michael. \emph{The Audit Society: Rituals of Verification}. Oxford: Oxford University Press, 1997.

Raji, Inioluwa Deborah, Emily M. Bender, Amandalynne Paullada, Emily Denton, and Alex Hanna. ``AI and the Everything in the Whole Wide World Benchmark.'' arXiv:2111.15366 (2021).

Schlangen, David. ``Dialogue Games for Benchmarking Language Understanding: Motivation, Taxonomy, Strategy.'' arXiv:2304.07007 (2023).

Rashkin, Hannah, Vitaly Nikolaev, Matthew Lamm, Lora Aroyo, Michael Collins, Dipanjan Das, Slav Petrov, Gaurav Singh Tomar, Iulia Turc, and David Reitter. ``Measuring Attribution in Natural Language Generation Models.'' \emph{Computational Linguistics} 49, no. 4 (2023): 777--840. DOI: 10.1162/coli\_a\_00486.

Ru, Dongyu, Lin Qiu, Xiangkun Hu, Tianhang Zhang, Peng Shi, Shuaichen Chang, Jiayang Cheng, Cunxiang Wang, Shichao Sun, Huanyu Li, Zizhao Zhang, Binjie Wang, Jiarong Jiang, Tong He, Zhiguo Wang, Pengfei Liu, Yue Zhang, and Zheng Zhang. ``RAGChecker: A Fine-grained Framework for Diagnosing Retrieval-Augmented Generation.'' arXiv:2408.08067 (2024).

Samarinas, Chris, Alexander Krubner, Alireza Salemi, Youngwoo Kim, and Hamed Zamani. ``Beyond Factual Accuracy: Evaluating Coverage of Diverse Factual Information in Long-form Text Generation.'' In \emph{Findings of the Association for Computational Linguistics: ACL 2025}, 13468--13482. 2025. DOI: 10.18653/v1/2025.findings-acl.693.

Sharma, Aman. ``Proxy-Based Evaluation and the Limits of Assurance in AI Governance.'' \emph{ACM AI Letters} (2026). DOI: 10.1145/3828672.

Sharma, Manasi, Chen Bo Calvin Zhang, Chaithanya Bandi, Clinton Wang, Ankit Aich, Huy Nghiem, Tahseen Rabbani, Ye Htet, Brian Jang, Sumana Basu, Aishwarya Balwani, Denis Peskoff, Marcos Ayestaran, Sean M. Hendryx, Brad Kenstler, and Bing Liu. ``ResearchRubrics: A Benchmark of Prompts and Rubrics for Evaluating Deep Research Agents.'' arXiv:2511.07685 (2025).

Skinner, Quentin. ``Meaning and Understanding in the History of Ideas.'' \emph{History and Theory} 8, no. 1 (1969): 3--53. https://www.jstor.org/stable/2504188.

Song, Yixiao, Yekyung Kim, and Mohit Iyyer. ``VeriScore: Evaluating the Factuality of Verifiable Claims in Long-form Text Generation.'' In \emph{Findings of the Association for Computational Linguistics: EMNLP 2024}, 9447--9474. 2024. DOI: 10.18653/v1/2024.findings-emnlp.552.

Wei, Jerry, Chengrun Yang, Xinying Song, Yifeng Lu, Nathan Hu, Jie Huang, Dustin Tran, Daiyi Peng, Ruibo Liu, Da Huang, Cosmo Du, and Quoc V. Le. ``Long-form Factuality in Large Language Models.'' arXiv:2403.18802 (2024).

Zhao, Jujia, Zhaoxin Huan, Zihan Wang, Xiaolu Zhang, Jun Zhou, Suzan Verberne, and Zhaochun Ren. ``ReportLogic: Evaluating Logical Quality in Deep Research Reports.'' In \emph{Proceedings of the 64th Annual Meeting of the Association for Computational Linguistics (Volume 1: Long Papers)}, 8470--8502. 2026. DOI: 10.18653/v1/2026.acl-long.384.
\end{document}